\documentclass[proceedings]{rmaa}

\usepackage{rmaacite}

\renewcommand{\P}[1]{%
\ifnum#1=1\hbox{OW~168--326E}\fi
\ifnum#1=2\hbox{OW~167--317}\fi
\ifnum#1=3\hbox{OW~163--317}\fi
\ifnum#1=5\hbox{OW~158--323}\fi
\ifnum#1=0\hbox{OW~171--334}\fi}
\input psfig.tex
\title{Extreme Type IIp Supernovae as Yardsticks for Cosmology}
\author{P. H\"oflich$^{1}$, O. Straniero$^{2}$, M. Limongi $^{3}$, I. Dominguez $^{4}$ \& A. Chieffi $^{3}$}
\altaffiltext{1}{Dept. of Astronomy, University of Texas, Austin, TX 78681, USA}
\altaffiltext{2}{Osservatorio Astronomico di Collurania, 64100 Teramo, Italy}
\altaffiltext{3}{Osservatorio Astronomico di Roma, Via Frascati 33, 00040  Monteporzio,  Italy}
\altaffiltext{4}{University of Granada, Granada, Spain}

\fulladdresses{
\item P. H\"oflich: Dept. of Astronomy, University of Texas, Austin, TX 78712, USA
                        (pah@hej1.as.utexas.edu).}

\shortauthor{H\"oflich et al.}
\shorttitle{Extreme SN~IIp and Cosmology}


\abstract{
Evolutionary effects with redshift of core collapse supernovae and their application
to cosmology have been studied based on an extensive grid of stellar models 
 between 13 and 25  $M_\odot$, and their light curves after the explosion.

With decreasing metallicity Z and increasing mass, progenitors tend to explode as compact
Blue Supergiants (BSG) and produce subluminous supernovae which are about $1.5^m$ dimmer
compared to 'normal' SNeII  with Red Supergiant (RSG) progenitors.
Progenitors with small masses tend to explode as RSGs even with low Z.
The consequences are obvious for probing the chemical evolution.

 We identify {\it extreme SNe~IIp} supernovae as a rather homogeneous class which may  allow
their use as standard candles for distance determination accurate within $30 \%$.
 Due to their unique light curves, no spectra need to be taken for their identification, and 
follow-up observations can be limited to a very
small dynamical range in brightness.
 This means that distance determination by  'extreme SNe~IIp" are possible up to redshifts
of $\approx 3 $ using 8-meter class telescopes.  SIRTIF may push the limit by another magnitude.
  }
\resumen{
Hemos estudiado los efectos evolutivos en las 
supernovas gravitacionales debidos al {\it redshift} y su aplicaci\'on en 
cosmolog\'\i a bas\'andonos en una extensa red de modelos con masas 
comprendidas entre 13 y 25 M$_\odot$ y sus curvas de luz.  

Al disminuir la metalicidad, Z, e incrementar la masa, las explosiones
tienden a producirse cuando el progenitor es una supergigante azul (BSG) 
 y consecuentemente se obtienen supernovas subluminosas, $\approx 1.5^m$ m\'as
d\'ebiles que una supernova {\it normal} producida en la explosi\'on de 
una supergigante roja (RSG). Los progenitores de baja masa tienden a explotar 
como supergigantes rojas incluso cuando Z es peque\~na. Analizamos las consecuencias 
que de ello se derivan en la evoluci\'on qu\'\i mica. 
Obtenemos que las {\it SNe IIP extremas} constituyen una clase muy homog\'enea, 
por lo que podr\'\i an emplearse como indicadores de distancia con una 
 fiabilidad de un 30$\%$. Debido a que presentan una curva de luz muy 
 caracter\'\i stica, pueden identificarse sin recurrir al espectro, y 
 su seguimiento observacional se limitar\'\i a un reducido intervalo de
 brillo. 
Esto hace posible el empleo de las {\it SNe IIP extremas} como 
indicadores de distancia hasta un {\it redshift} $\approx$ 3 usando 
telescopios de clase 8 metros. SIRTIF podr\'\i a aumentar el l\'\i mite 
observacional en 1 magnitud.

}

\begin{document}

\maketitle

\section{Introduction}
 Supernovae are among the brightest single objects which may reach the same brightness as
the entire host galaxy. This allows to measure distances on cosmological scales if the
intrinsic brightness of the object is known.
During the last few years, the main attention has been drawn by
Type Ia Supernovae because the homogenity in their properties  allows accurate
distance determinations based on either  statistical correlations in combination with exact
calibrations by $\delta-Ceph.$ stars  (Phillips 1993, Saha et al. 1997), or
detailed models. Both the empirical  and the theoretical approach provided consistent
values for $H_o$ (e.g. M\"uller \& H\"oflich 1994, Ries et al. 1995,
 Hamuy et al. 1996, H\"oflich \& Khokhlov 1996, 
 Nugent et al. 1996).
The routine detections of SNe~Ia at redshifts of 0.5 to 1.2
 provided strong evidence for a
positive cosmological constant (e.g. Perlmutter et al. 1999, Riess et al. 1999).
 For the latter results, an absolute accuracy of about 10 \% is required.
This leaves potential systematic effects of SNe~Ia properties with redshift
as major concern (H\"oflich et al. 1998), and it explains the goal to extend the
current efforts to large redshifts. In the current scenario, SNe~Ia are thermonuclear
explosions of white dwarfs which have grown to the Chandrasekhar mass $M_{Ch}$
 by accretion of H or He rich material  from a companion, and by  burning of the accreted 
 material to C/O. The strength of He shell flashes and the wind around accreting WD is 
very metal dependent. For Z  $\lesssim 0.1 \times $solar, the WD may not reach $M_{Ch}$
  (Nomoto et al. 2000, fully consistent with our findings).
 SNe~Ia may not (or rarely) occur at large redshifts.
From the observational point of view,
another restriction for  the use of SNe~Ia is that both spectra and light curves have to be taken to identify the
objects. 

The other class of SN,
core collapse supernovae, are thought to be the final results of massive stellar evolution for stars with main sequence masses
$\gtrsim 10 M_\odot $.
The light curves and spectra depend sensitively on the initial stellar mass, 
metallicity, mass loss and explosion energy.
 Therefore, these objects show a wide variety of brightness and properties of the light curves which prevents their use as quasi standard
candles. On the other hand, these objects  will occur soon after the initial star formation period and, therefore, can be used
to probe the structure of the universe at high z.

 In this work, we present a study focussed on core collapse supernovae to answer the following questions:
How do the light curves of core collapse supernovae depend on the metallicity which must be expected to decrease with redshift?
Can we identify a subclass among the core collapse supernovae which can be used as quasi-standard candles, and what accuracy do we
expect? Can this subclass be identified purely by their light curves, without a follow-up which
requires to "go" much fainter than maximum light?
\begin{figure}
  \begin{center}
    \leavevmode
    \psfig{figure=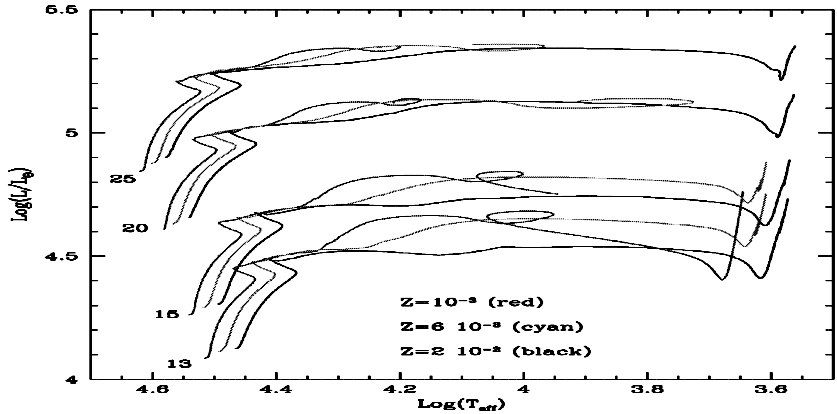,width=12cm,rwidth=13.3cm,clip=,angle=270}
      \vskip -0.3cm
    \caption{Stellar evolution from the main sequence to the onset of the core collapse
     for masses between 13 and 25 $M_\odot$
      with metallicities Z=0.02 (black), 0.066 (light grey) and 0.001(grey).}
    \label{fig:cartoon}
  \end{center}
      \vskip -0.4cm
\end{figure}
\section{Model Calculations}
 The stellar evolution has been calculated from the pre-main sequence to the onset
of the core collapse using the evolutionary code FRANEC
 (Chieffi \& Straniero 1989, Straniero et al. 1997 \& Chieffi et al. (1998).
 Stellar evolution models
have been constructed for masses of 13, 15, 20  and 25
$M_\odot$ and for  metallicities Z of 0.02, 0.006, 0.001, and 0. {\it Extreme SNe~IIp}  with
plateau phases longer than 50 to 60 days are
produced if most of the hydrogen rich envelope is retained (Fig. 1).
 Rather moderate mass loss does not alter
significantly the structure of both the core and the envelope of such stars and, consequently,
 the brightness of the LCs during the plateau phase (Chieffi et al. 2000 and below).
 For some of the model, mixing of
radioactive $^{56}Ni$ has been imposed. The model parameters have
been selected to demonstrate certain effects and to cover the  extremes rather than to
provide a 'best' model optimized to reproduce a given observation.
\begin{figure}
  \begin{center}
    \leavevmode
    \psfig{figure=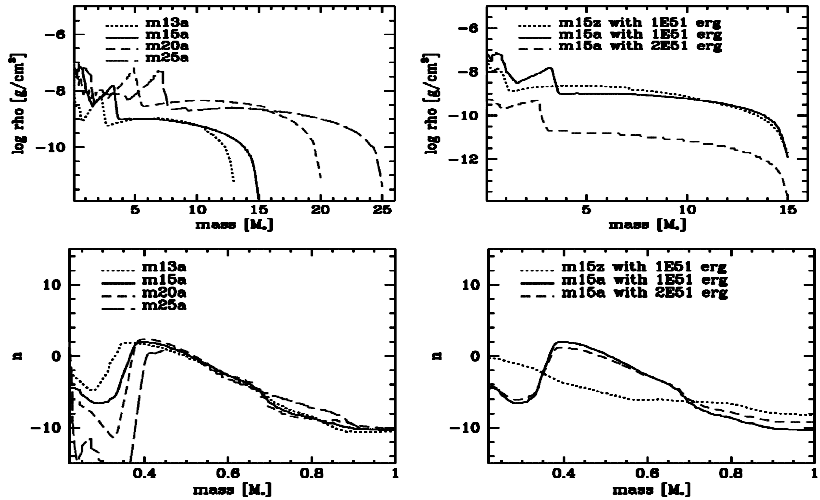,width=12.2cm,rwidth=13.5cm,clip=,angle=270}
      \vskip -0.3cm
    \caption{ Density profiles and density gradients n ($\rho \propto r^{-n}$)
     at day 5 for various masses, metallicities and explosion
     energies. The models are identified by their names {\rm mXXY} where XX gives the mass in solar
units and Y=a,z stands for solar and zero metallicity, respectively.
 Note the similarity of the density structures for all RSGs independent from mass and final
kinetic energy of the expanding envelope $E_{kin}$.
     }
    \label{fig:cartoon}
  \end{center}
      \vskip -0.2cm
\end{figure}
 For low metallicities Z, models explode  as  compact BSG
  ($R_\star \leq 100 R_\odot$) rather than as extended RSG ($500 R_\odot \leq  R_\star \leq 1500 R_\odot$).
We find that all the zero metallicity models end up as a BSG while all the solar
metallicity ones end up as a RSG. At intermediate Z, there is the general trend that
 the more massive stars end up as BSG while the less massive ones end up as RSG. The limiting
mass depends on Z (Fig. 1).
  Whether a star ends as a RSG or a BSG depends sensitively on the H-shell
burning or, more precisely, on the inner boundary of the H-rich layers which, in term, depends
on the chemical mixing of H/He assumed in the calculations which may change for a variety of reasons
such as common envelope evolutions, stellar rotation, turbulent mixing etc.

 Based on the final stellar structures, the hydrodynamical explosion and light curves
have been constructed using our one-dimensional radiation-hydro code (e.g. H\"oflich \& Khokhlov 1996,
and references therein). After the core collapse and the formation of the neutron star,
the explosion is triggered by depositing the explosion energy above the neutron star. The explosion
energy is adjusted to provide a final kinetic energy $E_{kin}$ of $1$ or $2 \times 10^{51}$erg.
 Due to the similarity in the final stellar structures, the density slopes
of the hydrogen rich envelopes are very similar for RSG progenitors during the phase of homologous
expansion (Fig.2).
Three phases can be distinguished for the light curves (Figs. 3 \& 4): 1)
  Most of the envelope is ionized. This phase
depends sensitively on the explosion energy, mixing of radioactive Ni, and the mass of the progenitor,
 e.g. either strong mixing or $E_{kin}\leq 1 foe $ will cause a steep and steady increase in B and V;
 2) The emitted energy is determined by the receding (in mass)
  of the H recombination front which is responsible for both the release of stored, thermal and
the recombination energy. At the recombination front, the opacity drops by about 3 orders
of magnitude providing a self regulating mechanism. If too little energy is released, the opacity
drops fast causing an increase in the speed at which the  the photosphere is receding.
 In term, this causes a larger energy release and vs. . Hydrogen recombines at a
specific temperature at or just below the photosphere. Consequently, the effective temperature and
the color indices remain largely unchanged during the recombination phase. 
Due to the flat density profiles of the expanding envelopes in the RSG case,
 the photospheric radius and, thus, the luminosity L stays almost constant.
 In contrast, for exploding BSG, the resulting steep density gradients
result in a steadily increasing radius and, since the recombination temperature hardly changes, in
a steadily increasing brightness (Fig. 4).  After the recombination front has
passed through the H-rich envelope,  the brightness drops fast.
During phase 3), L is given by the instant energy release by radioactive decay of $^{56} Co$.
  In all models,
the escape probability for $\gamma$ -rays remains very small up to 300 or 400 days after the
 explosion.  The size of the drop in L depends mainly on the amount of ejected $^{56}Ni$.

\begin{figure}
  \begin{center}
    \leavevmode
    \psfig{figure=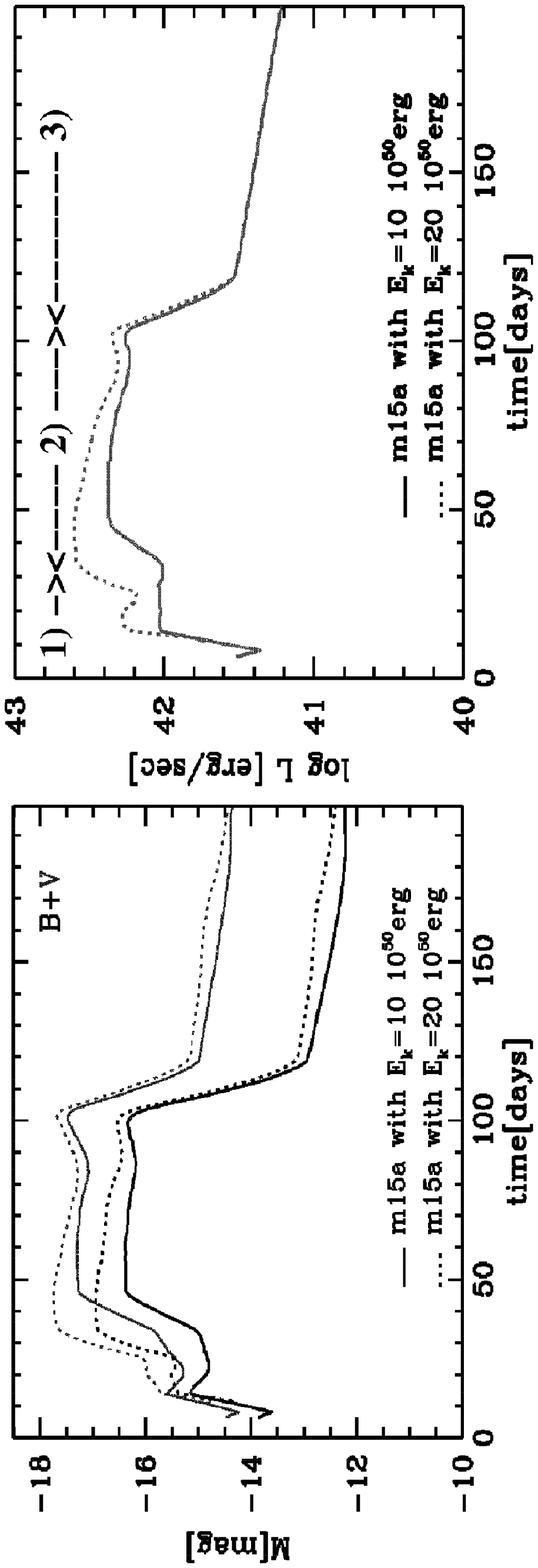,width=12.9cm,rwidth=14.2cm,clip=,angle=270}
      \vskip -0.3cm
    \caption{ Influence of the explosion energy on the bolometric, B and V light curves
 for a RSG of 15 solar masses without mixing of $^{56}Ni$.
 }
  \end{center}
  \begin{center}
     \leavevmode
    \psfig{figure=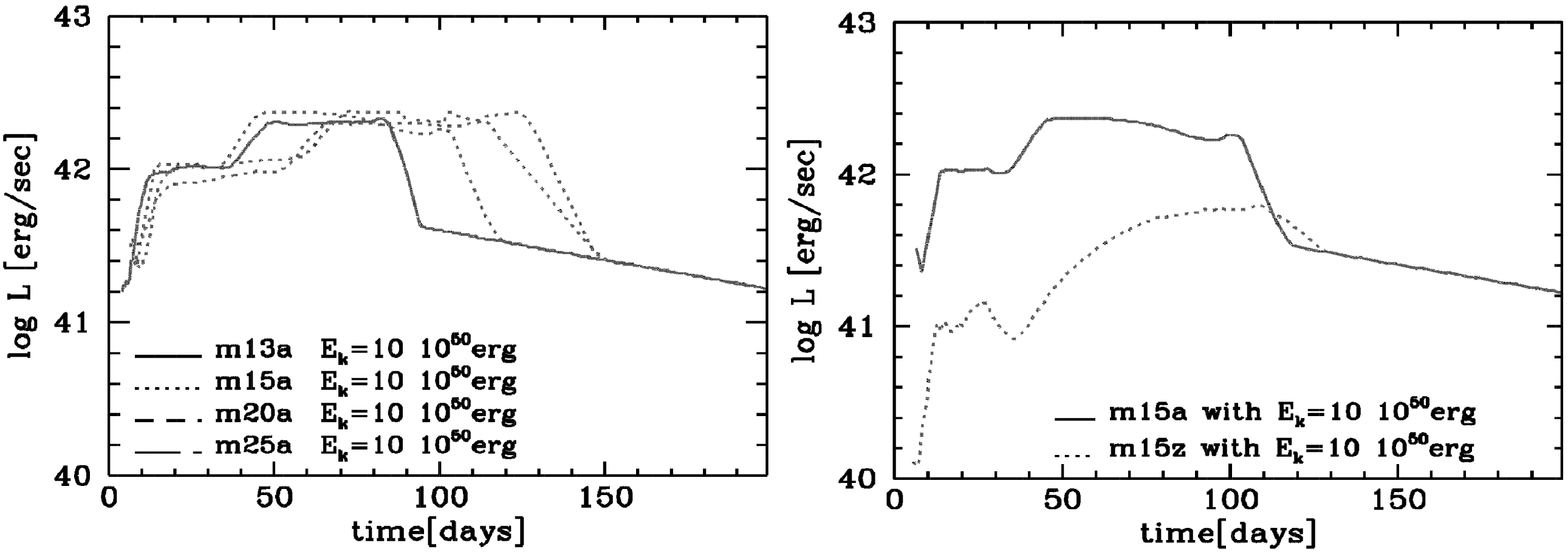,width=12.9cm,rwidth=14.2cm,clip=,angle=270}
    \caption{ Bolometric light curves for various masses and solar metallicity (left) and
     a 15 $M_\odot$ star with zero and solar metallicity (right) with $E_{kin} = 1E51 erg$.}
  \end{center}
   \vskip -0.2cm
\end{figure}

\section{Final Discussion and Conclusions}
 
Light curves for plateau supernovae have been studied. A set of
detailed calculations for stellar evolution  and light curves
have been computed  for a variety of initial masses, explosion
energies, mixing during the stellar evolution or during the
explosion, and metal abundances. 

 Based on our models,
we suggest the use of a subclass of Type II Supernovae, the {\it extreme SNeIIp}, as quasi
standard candles. These objects are characterized by a plateau phase in excess of 50 to
60 days (e.g. SN1999em).
 They can be understood as explosions of Red Supergiants which have undergone rather moderate
mass loss during the stellar evolution.
 The V brightness during the plateau phase changes/declines by about 0.2 to 0.7 $^m$.
The mean absolute brightness in V ($\approx 17.4 - 17.8^m$)
during the plateau phase is rather insensitive 
to the mass of the progenitor and the explosion energy  (within $\approx 0.6 ^m$).
 Note that line blocking in B and, in particular, in the  UV depends on the metallicity  causing a
somewhat larger spread.
The overall similarity of the LCs is caused by the similarity  of the density structures
of red giants, the resulting {\sl flat} density slopes,  the expanding H-rich envelope
 and the 'self-regulating' propagation of the recombination front which
determines the brightness during the plateau phase.

 In contrast, if the progenitor explodes as a blue supergiant, the resulting steep density profile
results in a long lasting phase of increasing photospheric radius and brightness. The maximum brightness is
 lower by  about $1.5^m$ compared to the explosion of a RSG because of the increased expansion work for BSGs.

 It is well known from SN1987A that low metallicity stars may explode as blue supergiant.
Qualitatively, this tendency is reproduced by our models.
 Note,  however, that  we showed above that, at the lower end of the
mass scale, the star may explode as a RSG even for Z as low as a 1.E-3.
  The mass dependence of the final outcome has  two main consequences. Firstly, the discovery
probability for SNe~II at high z will decrease with the progenitor mass. 
 The supernovae statistics will be systematically biased, starting at $z \approx 1$.
  The consequences for the study of the chemical evolution and the element production at high red-shifts
(e.g. by NGST) shall be noted.
Secondly, even at high redshifts, some {\it extreme SNe~IIp} should be visible. Taking their unique
properties, they may prove to be the key for the use of SN for cosmology at high z before
SNe~Ia occur.

 Although the  use of {\it extreme SNe~IIp} will not achieve the same accuracy as Type Ia Supernovae, there are some
distinct advantages: 1) due to their unique light curves and colors, no spectrum is required for
identification. 2) The requirements on the time coverage of the light curves are very moderate:
 Three or four deep images with a sample rate of 50 to 60 days in the rest frame
  will allow their discovery, 
 identification and their use for cosmology. At some time, two color images should be taken to
deselect flare star and to get at handle on the reddening.
  3) Finally, there is no need to follow the light curves
after the plateau towards dimmer magnitudes. For the use of SNe~Ia, the requirement to obtain a
spectrum limits the current use of SNIa  of $\approx 24^m$ if 8m-class telescopes are employed.
For the {\it extreme SN~IIp},  1) to 3) implies that the largest ground based telescopes 
 with IR detectors can  be used as search instruments which
pushes the limit to about $27 $ to $28^m$. Therefore, {\it extreme SNe~IIp} may be used up to $z \approx 3$
 using 8-meter class telescopes.
  SIRTIF may push the limit by another magnitude by long time exposures.

One potential pitfall is the unisotropic luminosity caused by aspherical explosions of core collapse SN.
 In general, the light of core collapse supernovae is polarized by $\approx  0.5... 1 \%  $
(e.g. Wang et al.  2000). Polarization of this size corresponds to asymmetries in the envelope which
produce directional dependence in the    observed L of $\approx 0.3 ~...~0.6^m$   (H\"oflich, 1991).
 However, extended H rich envelopes tend to spherize the H-rich layers of the
  envelopes even if the explosions are assumed
jet-like (Khokhlov, H\"oflich \& Wang, 2000 in preparation). This tendency is consistent with recent
observations for SN1999em (Wang 2000, private communication).
 
The statistical data base for {\it extreme SNe~IIp} is very incomplete. For the years 1998 and 1999,
 about 5 to 10 \% of all nearby SNII fall into this category  making the rate for this type
about a factor of 3 to 5 less abundant than SNe~Ia. However, the star formation rate at redshifts between     
2 and 3 was higher by a factor of $\approx 3 ... 5 $ (Kravtsov \& Yepes 2000)
 compared to the current rate, making the expected rates comparable to those of SNe~Ia. 
 For more details, see Chieffi et al. 2000.

\acknowledgements 
\noindent
{\it Acknowledgments:}
PAH would like to thank the people at the Observatories in Teramo
and Monteporzio for the hospitality during his stay when most of 
the work has been done.
This research was supported in part by  NASA Grant LSTA-98-022.
 The calculations for the explosion and light curves were done on a cluster of workstations 
financed by the John W. Cox-Fund  of the Department of Astronomy at the 
University of Texas, and processors donated by AMD.


\end{document}